\documentclass{article}
\usepackage{graphicx}
\usepackage{color}
\usepackage{amssymb,amsmath}

\newcommand{\revis}[1]{#1}

\begin{document}

\title{Ground-state properties of tubelike\break{} flexible polymers}

\author{Thomas Vogel$^{1,3,}$\thanks{Corresponding author: t.vogel@fz-juelich.de}\;,
Thomas Neuhaus$^{2,}$\thanks{t.neuhaus@fz-juelich.de}\;,\\
Michael Bachmann$^{1,3,}$\thanks{m.bachmann@fz-juelich.de}\;,
and Wolfhard Janke$^{1,}$\thanks{janke@itp.uni-leipzig.de}\\[.5cm]
{\normalsize{$^{1}$ Institut f\"ur Theoretische Physik and Centre for Theoretical Sciences (NTZ),}}\\
{\normalsize{Universit\"at Leipzig, Postfach 100\,920, 04009 Leipzig, Germany}}\\
{\normalsize{$^{2}$ J\"ulich Supercomputing Centre (JSC), Forschungszentrum J\"ulich,}}\\
{\normalsize{52425 J\"ulich, Germany}}\\
{\normalsize{$^{3}$ Institut f\"ur Festk\"orperforschung, Theorie II (IFF-2),}}\\
{\normalsize{Forschungszentrum J\"ulich, 52425 J\"ulich, Germany}}}


\date{}

\maketitle

\begin{abstract}
\noindent
In this work we investigate structural properties of native
states of a simple model for short flexible homopolymers,
where the steric influence of monomeric side chains is
effectively introduced by a thickness constraint. This
geometric constraint is implemented through the concept of
the global radius of curvature and affects the
conformational topology of ground-state structures.  A
systematic analysis allows for a thickness-dependent
classification of the dominant ground-state topologies. It
turns out that helical structures, strands, rings, and coils
are natural, intrinsic geometries of such tubelike objects.\medskip
\end{abstract}

\section{\label{sec:intro}Introduction}

The structural properties of macromolecules composed of
covalently bound atomic complexes are of major interest as
the functionality of these molecules, e.g., biopolymers,
strongly depends on the formation of stable ground-state
conformations with several substructures. One can resolve
these structures by means of several experimental techniques
like molecular microscopy, NMR or X-ray analyses.

As these techniques are relatively costly and can often
hardly be generalized, the structural behavior of polymers
and its modeling got into the focus of computer simulations,
too. In early approaches, polymers were modeled as
topologically one-dimensional strongly coarse-grained
bead-spring systems, which were treated by means of Monte
Carlo and molecular dynamics computer
simulations~\cite{rosen55jcp,landbind05buch,frenksmit02buch}.
Later, all-atom peptide simulations were performed with the
intention to study real biopolymers, in particular their
native states and the folding pathway to native
states~\cite{gnanarev03cosb,janke08lnp}.  As this is still a
great challenge and moreover restricted to comparatively
short objects, the interest in the earlier simple models
continues until today. There is still a great variety of
one-dimensional stringlike models of flexible polymers under
current investigation (see,
e.g.~\cite{hsu07epl,rpb07pre,schna07prl,schna09cpl}).

Biopolymers do have side chains, however, which are
responsible for the variety of forms and functions. As these
side chains involve strict steric constraints for the
structure of the polymer, the central question is therefore:
What degree of abstraction (coarse-graining) is reasonable
to treat certain features of real (bio)polymers or, the
other way around, which features can we reliably study with
a certain degree of abstraction? One can, for example,
understand certain universal properties of the well-known
coil-globule transition by studying the simplest stringlike
models~\cite{grassb97pre,voge07pre}, whereas the formation
of secondary structures will generally not be answered
satisfactorily.\footnote{We refer here mainly to ground
states of models for flexible polymers \revis{with a single
length scale}. Of course, there are studies of simple
polymer models dealing with the problem of secondary
structure formation, where helices were found for example as
(long-living) transient states during the folding process,
as ground states for stiff polymers\revis{, or at special
ratios of different length scales in the
system~\cite{kemp98prl,noguchi98jcp,magee06prl,sabeur08pre}.}}

We are going to approach this problem in the present work by
studying in detail some kind of ``intermediate'' model. It
is derived from a linelike model where additionally the
steric influence of monomeric side chains in real
biopolymers will be effectively introduced, without taking
into account further microscopical details, by a geometric
thickness constraint, extending the model to a
three-dimensional
tube~\cite{banamarit03jpcm,banamarit03rmp,neuh06}.  An
elegant possibility to implement the thickness constraint is
provided by the concept of the global radius of curvature
which ``provides a concise characterization of the thickness
of a curve''~\cite{gonzmad99pnas}. We will see, that we
already get quite interesting qualitative and quantitative
answers to the question of secondary structure formation
under these conditions.\footnote{There are recent,
interesting studies of a tube model, showing that helices
can form entropically driven. In these studies, however,
solvent particles and solvent effects are explicitly
introduced and play an important
role~\cite{snir05sci,snir07pre,hansgo07prl}.}

The structure of this paper is as follows: In
Sect.~\ref{sec:models} we first introduce the model we use
and shortly describe our methods, and in Sect.~\ref{sec:grc}
we explain in more detail, how the thickness is implemented
in our model. After introducing briefly some observables in
Sect.~\ref{sec:obs}, we study in Sect.~\ref{sec:A} the
ground states of the model in great detail, depending on the
thickness with inter-monomer interaction parameters kept
fixed. Furthermore, in Sect.~\ref{sec:selected} we look at
special problems like crystallization into regular lattice
structures and the appearance of biologically relevant
structures depending on a variable interaction length
scale. The paper concludes in Sect.~\ref{sec:summary} with a
summary of our main findings.

\section{\label{sec:Techni}Technical details}
\subsection{\label{sec:models}Model and Methods}

To model flexible homopolymers we use a thick, tubelike
off-lattice chain with fixed bond length and pure
Lennard--Jones (LJ) interaction between all, except for the
neighboring, monomers:
\begin{equation}
V_{\mathrm{LJ}}(r_{ij})=
4\left(\left(\frac{\sigma}{r_{ij}}\right)^{12}
-\left(\frac{\sigma}{r_{ij}}\right)^6\right)\,,
\end{equation}
where $r_{ij}=|\mathbf{x}_i-\mathbf{x}_j|$ is the distance
between two monomers at positions $\mathbf{x}_i$ and
$\mathbf{x}_j$.  Note that $V_{\mathrm{LJ}}(r_{ij})=0$ for
$r_{ij}=\sigma$ and that the potential minimum
$V_{\mathrm{LJ}}(r_{ij})=-1$ is located at $r_{ij}^{\rm
min}=2^{1/6}\,\sigma$. The bond length $r_{i,i+1}$ is set
to $1$.

We investigate the model by means of Monte Carlo computer
simulations and numerical methods. For the ground-state
search, one may use for example the multicanonical
method~\cite{bergneuh91plb,bergneuh92prl,bittner08prep}, the
efficient random walk algorithm introduced by Wang and
Landau~\cite{wangl01prl,zhoubhatt05condmat} or parallel
tempering~\cite{partemp1,partemp2}. For this purpose, all
methods work nearly equally well as one does not have to
care about the quality of the sampling of the whole
configuration space and the performance depends mainly on
the parameters of the methods. Additionally, we use standard
conjugate gradient methods to refine the
results~\cite{conjug_numer_rec}.
For the methodologically more challenging task of studying the
thermodynamic behavior~\cite{letter,partII}, we use the
multicanonical method, calculating the multicanonical weights
with the help of the Wang--Landau algorithm~\cite{bittner08prep}.
Chains with different thicknesses have always been simulated
independently from each other, not at least to avoid
uncontrollable correlations.

\subsection{\label{sec:grc}Thickness and Global Radius of Curvature}

What precisely motivates us to simulate coarse-grained
homopolymer models with geometric constraints like
``thickness''? And, how is this realized in the model?
Polymers in biology are not thin strings. Amino acids, and
thus proteins, do have rather extended side chains sterically avoiding each
other. It might therefore be useful to introduce a
constraint that mimics this volume exclusion.  Furthermore,
it has been shown~\cite{banamarit02prot} that tube models
for polymers allow for the formation of stable regions of
biologically relevant (sub)structures like helices or
sheets, in contrast to simpler linelike polymer
models~\cite{banagonz03jsp}. In addition, the introduction of
such a geometric constraint restricting the conformational
space might even lead to some technical advantages for
finding minimal energy conformations in sophisticated
stochastic ground-state searches~\cite{partII}.

To define the self-avoiding tube, we use the concept of the global
radius of curvature $r_\mathrm{gc}$, which is a concise characterization
of the thickness of a curve~\cite{gonzmad99pnas,banagonz03jsp}. The
global radius of curvature is defined as the minimal radius of all
circumcircles $r_\mathrm{c}$ of any three monomers in the chain:
\begin{equation}
r_\mathrm{gc}(\mathbf{X}):=\min\{r_\mathrm{c}(\mathbf{x}_i,\mathbf{x}_j,\mathbf{x}_k),\, \forall\, 1\leq i<j<k\leq N\}\,.
\end{equation}
The circumradius $r_\mathrm{c}$, i.e., the radius of curvature, of three
points located at positions $\mathbf{x}_i$, $\mathbf{x}_j$
and $\mathbf{x}_k$ can be calculated as
\begin{equation}
r_\mathrm{c}(\mathbf{x}_i,\mathbf{x}_j,\mathbf{x}_k)=\frac{r_{ij}\,r_{ik}\,r_{jk}}{4A_\Delta(\mathbf{x}_i,\mathbf{x}_j,\mathbf{x}_k)}\,,
\end{equation}
where $A_\Delta(\mathbf{x}_i,\mathbf{x}_j,\mathbf{x}_k)$ is
the area of the triangle spanned by the three points at
$\mathbf{x}_i$, $\mathbf{x}_j$ and $\mathbf{x}_k$. Note,
that the model uses a three-point interaction between
monomers, in contrast to two-point interactions typically
used in hard-sphere models to incorporate volume exclusion
effects.  Given a polymer conformation
$\mathbf{X}=(\mathbf{x}_1,\dots,\mathbf{x}_N)$ with $N$
mono\-mers, we then define as its thickness $d$ twice the
global radius of curvature $r_\mathrm{gc}(\mathbf{X})$:
$d(\mathbf{X})=2r_\mathrm{gc}(\mathbf{X})$.

The other way around, given a thickness constraint $\rho$
such that $d\geq2\rho$, one can construct an excluded volume
depending on $\rho$ around two monomers, which is
``forbidden'' for any other monomer. A polymer conformation
then complies with the thickness constraint if any other
monomer resides outside these circles.  The partition
function reads in this case:
\begin{equation}
\label{partfunc}
Z_\rho=\int \mathcal{D}X\Theta(r_\mathrm{gc}(\mathbf{X})-\rho)\,\mathrm{e}^{-E(\mathbf{X})/{T}}\,,
\end{equation}
where $\Theta(z)$ is the Heaviside function and
\begin{equation}
E(\mathbf{X})=\sum_{i,j>i+1}V_\mathrm{LJ}(r_{ij})
\end{equation}
is the energy of the conformation $\mathbf{X}$.

Intuitive illustrations of this approach can be found
in~\cite{neuh06}, where it has been chosen for the analysis
of ringlike tube polymers. It is used as well
in~\cite{letter,partII} where the pseudophases of secondary
structures for tubelike polymers are investigated. As a
remark: Even though some universal properties (such as, e.g., critical exponents) do not depend
on the exact definition of the thickness, 
the model may behave differently by imposing the hard constraint $d=2\rho$, i.e.,
by replacing $\Theta(r_\mathrm{gc}(\mathbf{X})-\rho)$ with
$\delta(r_\mathrm{gc}(\mathbf{X})-\rho)$ in the partition
function~(\ref{partfunc})~\cite{neuh06}.

\section{\label{sec:results}Results and Discussion}

In the following, we first analyze in detail the ground
states and ground-state regions of tubelike polymers for
chain lengths $8\leq N\leq13$ with Lennard--Jones parameter
$\sigma=1$, where $V_\mathrm{LJ}=0$ for covalently bound
monomers. Furthermore, we investigate the influence of the
tube thickness in connection with the length scale $\sigma$
of the nonbonded LJ interaction on the formation of
different classes of structures. In this analysis,
particular emphasis will be dedicated to secondary
structures such as helices and strands, being most relevant
for segments of biomolecules.

\subsection{Conformational Classification Observables}
\label{sec:obs}

We characterize and identify conformations by their energy
$E(\mathbf{X})$, but naturally also by geometrical properties
such as the end-to-end distance $r_{\mathrm{end}}$ or radius of
gyration $r_{\mathrm{gyr}}$. However, as this turns out to be not sufficient
to distinguish between all conformations, we also take
into account local radii of curvature
$r_{\mathrm{lc},i}:=r_{\mathrm{c}}(\mathbf{x}_i,\mathbf{x}_{i+1},\mathbf{x}_{i+2})$,
which are related to the bending angles between two bonds
$\vartheta$ via
$\vartheta=1/r_\mathrm{lc}+\mathcal{O}(r_\mathrm{lc}^{-2})$
for small $\vartheta$, as well as torsion
angles $\phi\in(-\pi,\pi]$.

We speak of a $\kappa_0$-conformation, if the chain has a constant
local curvature at all monomer positions,
i.e., $r_{\mathrm{lc},i}=\mathrm{const}\,,\forall
i$.  Analogously, we call a structure with
constant torsion angles a $\tau_0$-conformation~\cite{koch98jgg}. For example, a prominent structure
with both, $\kappa_0$- and $\tau_0$-property, is the perfect
$\alpha$-helix.
A conformation is called ``closed'' if the distance between
both ends of the chain, i.e., $r_\mathrm{end}$, resides in
the close vicinity of the Lennard--Jones minimum, whereas a
``symmetric'' structure exhibits a symmetry of the torsion
angles with respect to the center of the chain. Nice
examples for ``closed'' $\kappa_0$ conformations are twisted
circles of constant curvature (in German, so-called
``windschiefe Kreise''~\cite{cesaro26}). Finally, in
``flat'' conformations, the backbone has an almost
two-dimensional, planar structure, where all torsion angles
converge to 0, i.e., $\sum_i\phi_{\pi/2}^{(i)}\to0$, where
\begin{equation}\label{eq:t90}
\phi_{\pi/2}:=\min(|\phi|,\pi-|\phi|)\,.
\end{equation}

\subsection{Thickness Dependent Ground-State Properties\\ of Tubelike Polymers}
\label{sec:A}

For our comprising study of the ground states of the model,
we first set the Lennard--Jones parameter to $\sigma=1$. In
this case, the ``natural thickness''
$r_\mathrm{gc}^\mathrm{min}$, i.e., the global radius of
curvature of the ground state of a flexible LJ polymer
\emph{without} thickness constraint, is about half the interaction
length $r_{ij}^\mathrm{min}/2=2^{-5/6}\approx0.56$, which
thus sets a reasonable bound for the thickness
constraint.\footnote{Actually, due to the discrete nature of
the bead-stick polymer model, we measure a ``natural
thickness'' $\gtrsim0.59$ for all
polymers considered.} Below this value, the thickness
constraint does not influence ground-state properties at
all. Thus, in the following, we only consider tube polymers
with $\rho>r_\mathrm{gc}^\mathrm{min}$.

\begin{figure}[t]
\includegraphics[width=\textwidth]{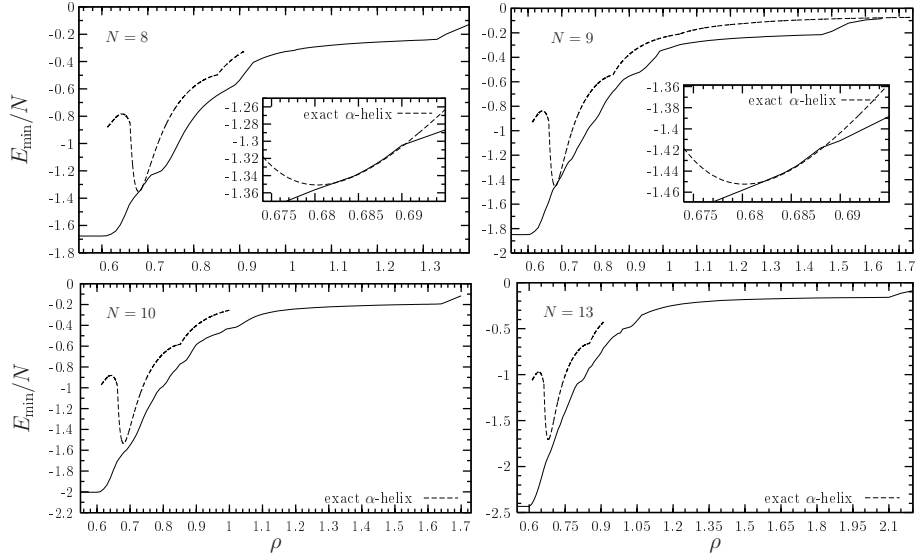}
\caption{Energies of ground states depending on the
thickness constraint $\rho$. \textit{Dashed lines} show for
comparison the energy for exact $\alpha$-helices.}
\label{fig:1}
\end{figure}
\begin{figure}
\includegraphics[width=.5\textwidth]{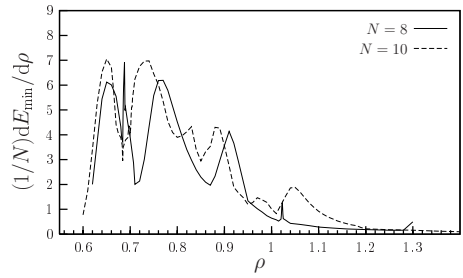}
\caption{Numerical derivatives of the energies in Fig.~\ref{fig:1} with respect to $\rho$ for $N=8$ and $10$.}\label{fig:1a}
\end{figure}

\enlargethispage{\baselineskip}
Figure \ref{fig:1} shows the energies of the ground states
for various chain lengths in dependence of $\rho$ in
comparison to the energy of the corresponding calculated
space-filling (perfect) $\alpha$-helix. This helix is
defined as the helix
$(x,y,z)=(r\sin\phi,r\cos\phi,p\phi/2\pi)$ with a pitch $p$
such that the surface of the tube has a~self-contact at the
cylinder with radius $r$ and the radius $r$ is minimal under
the thickness constraint. In other words, the optimally
packed, i.e., space-filling, $\alpha$-helix corresponds to the
transition between the two qualitatively different regimes
of $p/r>c^*$ and $p/r\ll1$~\cite{marit00nature}. The
computation of this helix is non-trivial, as the critical
ratio $c^*$ depends on the discretization level and ranges
from $c^*\approx2$ for $\rho\approx0.7$ to $c^*=2.512$ for
the continuous case, which is equivalent to
$\rho\to\infty$. An interesting and detailed
discussion of compact helix formation and the critical ratio
$c^*$ can be found also in~\cite{snir07pre}.

\begin{figure}[t]
\includegraphics[width=.8\textwidth]{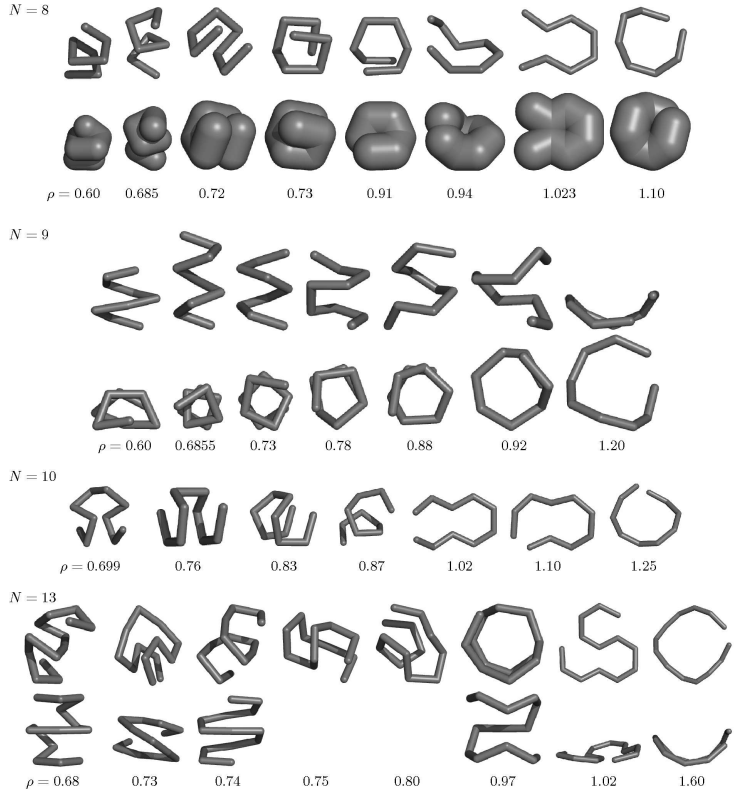}
\caption{Ground-state conformations for selected thickness
parameters $\rho$ for $N=8$, $9$, $10$ and $13$ (from \textit{top} to
\textit{bottom}). \revis{The second row for $N=8$ shows the same ground states with appropriate
thickness, to give a better idea of the ``real'' objects we
are investigating.} The second rows for $N=9$ and $N=13$
show an alternative view of the same configuration. For
reasons of better visibility, the thickness is not shown in
the proper scale \revis{(except for the second row)}.}\label{fig:2}
\end{figure}
\begin{figure}[t]
\includegraphics[width=\textwidth]{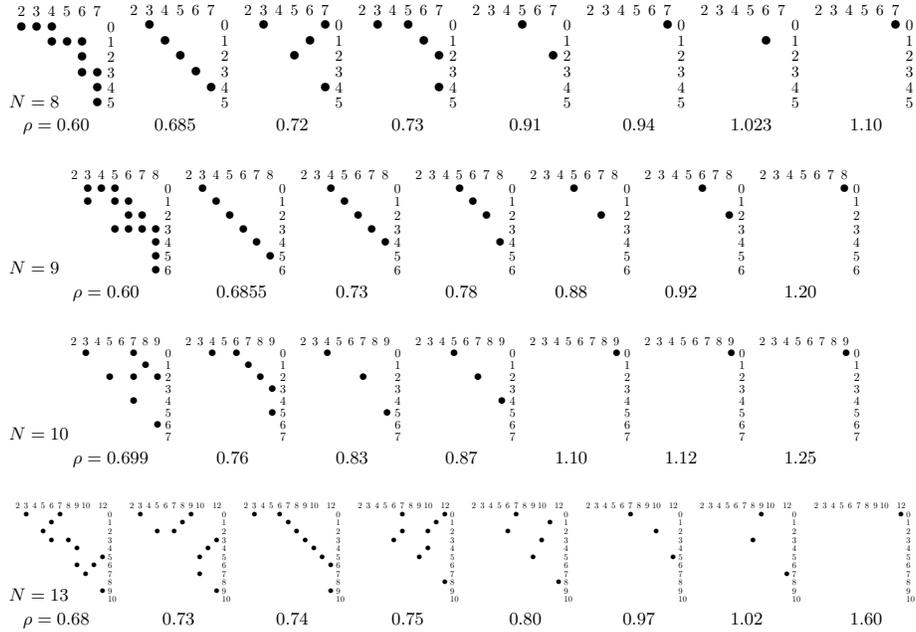}
\caption{Contact maps of the conformations shown in
Fig.~\ref{fig:2}. The axes show the monomer numbers $i$ and
$j$, the entries of the matrix are set ($\bullet$), if two
monomers $i$ and $j$ are in contact with each other, i.e.,
if the distance between them in the three-dimensional
structure is shorter than a certain threshold value. This
value is here slightly larger than the minimum distance of
the LJ potential $r_{ij}^\mathrm{min}$.}
\label{fig:2map}
\end{figure}
\begin{figure}
\includegraphics[width=\textwidth]{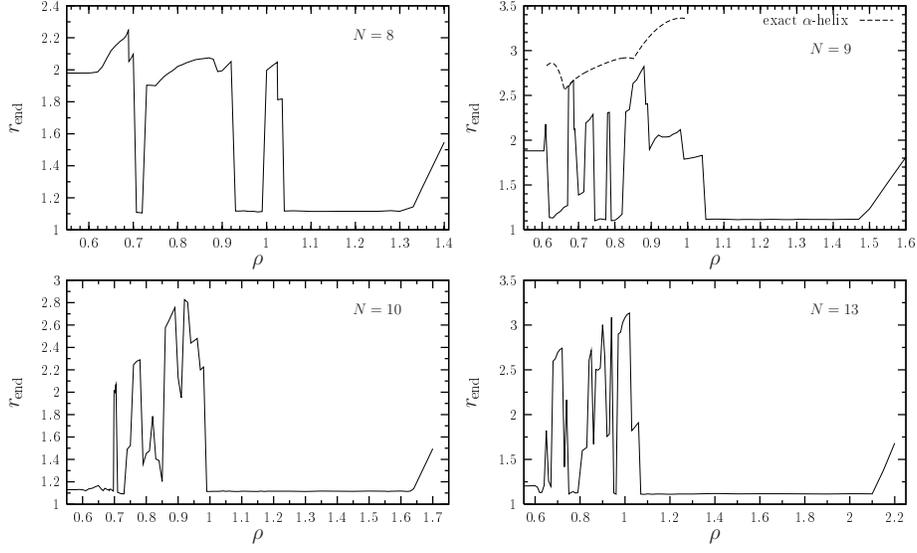}
\caption{End-to-end distances of ground-state conformations depending on the thickness constraint $\rho$.
The \textit{dashed line} for $N=9$ shows for comparison the
end-to-end distance for the exact $\alpha$-helix.}
\label{fig:3}
\end{figure}
\begin{figure}
\includegraphics[width=\textwidth]{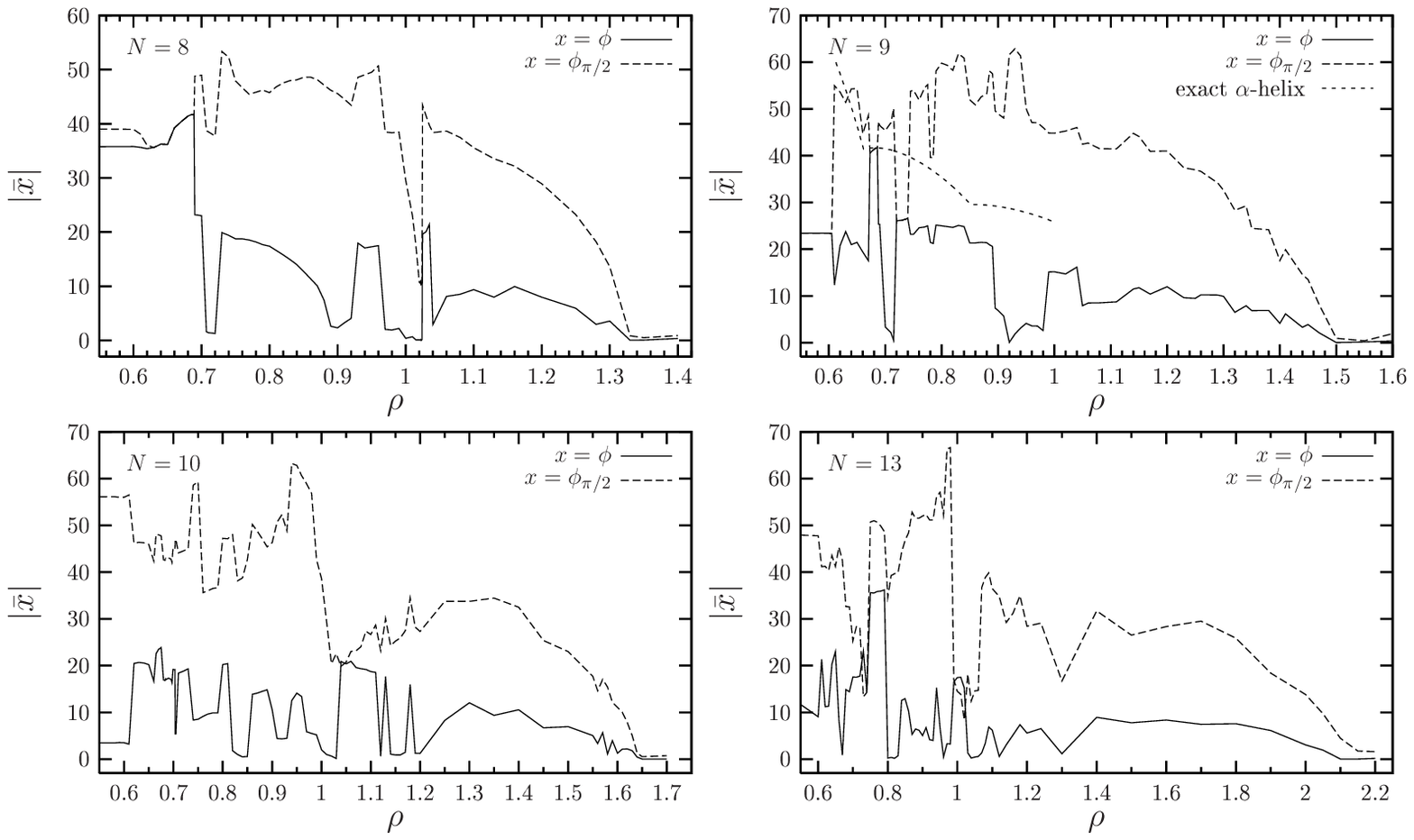}
\caption{Mean torsional angles of ground-state conformations 
depending on the thickness constraint $\rho$. Shown are
absolute mean values of $\phi$ and
$\phi_{\pi/2}:=\min(|\phi|,\pi-|\phi|)$. The
\textit{short-dashed line} for $N=9$ shows the behavior for the exact $\alpha$-helix.}
\label{fig:4}
\end{figure}
For two exemplified chain lengths, $N=8$ and $N=10$, we have
plotted in Fig.~\ref{fig:1a} the derivatives
$\mathrm{d}E/\mathrm{d}\rho$ in order to emphasize regions
of structural activity. In these regions, where the
derivative exhibits peaks, noticeable qualitative
conformational transitions occur. To describe and understand
these different classes of ground-state conformations, we
visualize in Fig.~\ref{fig:2} significant
structures and plot in Fig.~\ref{fig:2map} the contact maps,
where a contact is counted, if the distance between two
monomers $r_{ij}<r_{ij}^\mathrm{min}+\epsilon$. We set
$\epsilon\approx0.2$ but, of course, the contact maps do not
depend on minor variations of
$\epsilon$.\footnote{Furthermore, due to the small size of
the systems, the contact maps do not become more meaningful
by scaling $\epsilon$ with the thickness in some way,
instead of keeping $\epsilon$ constant.} In addition, in
Figs.~\ref{fig:3} and~\ref{fig:4} end-to-end distances and
mean torsional angles discussed above are shown. Based on
this data, we can classify the generic behavior by
introducing three general regions: thin, intermediate, and
thick tubes. Let us now look at these regions in more
detail.

\subsubsection{Thin tubes}
\label{sec:A_thin}

The thin-tubes region is, besides some singular points,
dominated by helical and helicallike conformations. We call
a conformation ``helical'', if $\bar\phi=\bar\phi_{\pi/2}$,
where $\bar x$ is the average along the chain, $\bar
x=(1/N)\sum_{i} x_i$, i.e., if all torsion angles lie in a
range $0\dots\,\pm\pi/2$ (cp. Eq.~(\ref{eq:t90}) and
Fig.~\ref{fig:4}; we do not distinguish between right- and
left-handed helices, but the sign must not change within the
conformation). Furthermore, the entries in the contact map
lie precisely parallel to the diagonal of the matrix in
these cases, a clear indication for helix structures (see,
for example, $N=8; \rho=0.685$ or $N=9;
\rho=0.73$ in Fig.~\ref{fig:2map}). ``Helicallike''
conformations share some properties with helical structures,
e.g., they exhibit a large, slightly increasing end-to-end
distance (cp. Fig.~\ref{fig:3}) with increasing thickness,
but the torsion-angle criterion above may be violated
(typically in a periodical manner) and the contact-map
entries do not form an exact parallel, but a line roughly
parallel to the diagonal (for example, at $N=8; \rho=0.6$ or
$N=9; \rho=0.78$). Generally, we find three interesting
effects looking at the contact maps in Fig.~\ref{fig:2map},
which have been mentioned above or will be discussed later
again: First we see, that polymer chains without thickness
constraint (see the maps for $N=8$ and $N=9$ with
$\rho=0.6$) do not have a pronounced structure. Just by
increasing the thickness a bit, clear helical structures
emerge, indicated by straight ``lines'' parallel to the
diagonal of the map (cp. $N=8$ and $N=9$ with
$\rho=0.685(5)$). Secondly, by increasing the thickness
further, we see, for example for $N=9$, that this parallel
``lines'' moves away from the diagonal, i.e., the helical
conformations are ``untwisting''. Finally, looking at the
contact maps for $N=13$, we see that tertiary effects come
into play, indicated by ``disrupting'' vertical ``lines'',
which is typically an indication for sheetlike structures.

Remarkably, within certain intervals ($N=8$: $0.63\leq
\rho\leq 0.688$; $N=9$: $0.673\leq \rho\leq0.6855$), the
ground-state conformations expand with increasing thickness
to a perfect space-filling helix with $\kappa_0$- and
$\tau_0$-property, i.e., an $\alpha$-helix with constant
bond- and torsion angles ($\bar{\phi^2}-\bar\phi\,\bar\phi$
and $\bar{r^2}_{\mathrm{lc}}-\bar r_{\mathrm{lc}}\bar
r_{\mathrm{lc}}$ vanish\footnote{A remark on the precision
of the simulation: The values of
$\bar{\phi^2}-\bar\phi\,\bar\phi$ and
$\bar{r^2}_{\mathrm{lc}}-\bar r_{\mathrm{lc}}\bar
r_{\mathrm{lc}}$ become even with the stochastic methods
smaller than $10^{-8}$ at this point, i.e., the difference
between any two torsion angles, for example, in the chain is
already less than $1.5\times10^{-4}\pi$.}). The comparison
of measured observables with the data for the exact
$\alpha$-helix is emphasized in the insets
of~Fig.~\ref{fig:1} and furthermore exemplarily shown in
Figs.~\ref{fig:3} and~\ref{fig:4}. We will resume the
discussion on this fact in Sect.~\ref{sec:alpha} below.

As a remark: What is the motivation to call these
conformations $\alpha$-helical in imitation of the real
biological $\alpha$-helix? In natural proteins,
$\alpha$-helices possess about $3.6$ amino acids per helix
turn~\cite{principles_short_eng} and have mainly constant
bond and torsion angles. If we construct a perfect space
filling helix with exactly 3.6 monomers per turn, we find
that it has a global radius of curvature of
$r_{\mathrm{gc}}\approx0.69$. Or, the other way around in
the region $0.6845\leq
\rho\leq 0.688$ (example for $N=8$), one counts
$3.576\ldots3.596$ monomers per turn, which is in very good
agreement with natural $\alpha$-helices.
We thus see the first biological relevant structure
realized by the simplest model with just Lennard--Jones
interaction and thickness but without any conformational
assumptions or additional input.

An above mentioned singular point is located in the vicinity
of the perfect helices at
$\rho\approx1/\sqrt{2}\approx0.71$, where ground states
attempt to crystallize in a regular simple cubic (sc)
lattice structure. We find for example for $N=8$ at
$\rho=0.73$ a $\kappa_0$-conformation almost fitting the sc
lattice (elsewhere called ``simple cubic lattice
helix''~\cite{dill90jcp}), which then untwists with
increasing thickness. We see the same tendency for longer
chains as well (see Fig.~\ref{fig:2} for visualizations and
Sect.~\ref{sec:alpha} for further discussion).  Note that a
perfect cube will not be a ground state at any thickness, as
the Lennard--Jones interaction length scale is larger than
the bond length. If we reconfigure the potential such that
its minimum value equals the bond length, i.e. set
$r_{ij}^{\rm min}=1$, we find indeed that the ground-state
conformations fit exactly into the simple cubic lattice
(i.e.~are exact cubes for adequate monomer numbers) up to
lengths of $\approx30$ monomers.  We will show this in more
detail in Sect.~\ref{cubes}.

At larger thickness we observe in Fig.~\ref{fig:2} extended
helicallike conformations, which may overlap due to the
shortness of the chains only at the end bonds.

\subsubsection{Intermediate tubes}

In the interval $0.9\lesssim \rho\lesssim1.0$, we observe an abrupt switch to
almost flat (cp. Fig.~\ref{fig:4}) and mostly closed
(cp. Fig.~\ref{fig:3}) conformations. One finds bended double-rings,
hairpins, and even conformations that are ``crystallized'' on a
two-dimensional honeycomb lattice (cp.~Fig.~\ref{fig:2}, $N=8$, $13$,
$\rho\approx1.02$). These curves are, of course, $\kappa_0$-curves as
well and have apparent similarities to $\beta$-sheets known from
secondary structures of biopolymers. We find in some small regions
competitions between mesomeric structures, i.e. structures with the
same monomer positions but different bond distributions (see, for
example, Fig.~\ref{fig:2}, $N=10$, $\rho\approx1.1$).

\subsubsection{Thick tubes}

At $\rho\approx1.1$, the ground-state conformation is (again)
``closed'' for all chain lengths. Here begins the region of
the twisted circles of constant curvature (``windschiefe
Kreise'')~\cite{cesaro26,koch98jgg}. With increasing
thickness, the rings become more and more flat until they
reach the two-di\-men\-sional ring at $\rho\approx N/2\pi$,
which is again a $\kappa_0$- and $\tau_0$-curve. Increasing
thickness just pushes apart the ring, what can clearly be
seen in the end-to-end distance and the torsion angles (see
Figs.~\ref{fig:3} and~\ref{fig:4}). For the somehow
pathological case of $\rho\to\infty$ one would reach the
limit of stiff rods.

\begin{figure}[b]
\includegraphics[width=.75\textwidth]{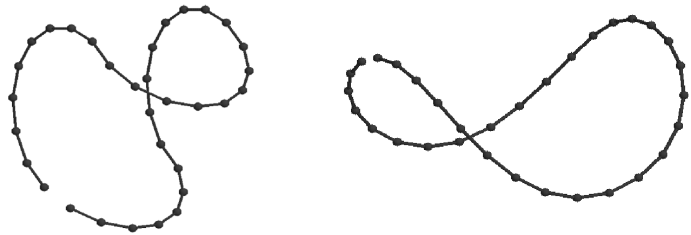}
\caption{Two examples for twisted circles of constant curvature $r_{\rm c}\approx2.6$ (\textit{left}) and $r_{\rm c}\approx3.6$ (\textit{right}) with $N=32$ monomers. See text for details.}
\label{fig:kreise}
\end{figure}

For purposes of illustration we display two examples from 
the class ``windschiefe Kreise''. The first kind consists
out of 4 half circles, which form a closed three-dimensional
curve. The left side of Fig.~\ref{fig:kreise} displays a
$N=32$ chain, which is a ground state of the theory and has
been obtained from simulations at $r_{\rm gc}=2.562915$ and
$r_{ij}^{\rm min}=1.6$. As can be seen each of the
half-circles consists of eight monomers. The second kind of
``windschiefe Kreise'' consists out of four helix sections,
that are joined together in such a way that the resulting
curve is closed again. The right side of
Fig.~\ref{fig:kreise} displays such a $N=32$ chain, which
was obtained from simulations at $r_{\rm gc}=3.624510$ and
$r_{ij}^{\rm min}=1.6$.

\subsubsection{General remarks}

\begin{figure}[h]
\includegraphics[width=.5\textwidth]{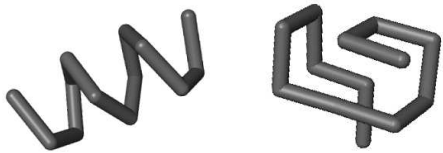}
\caption{Two $N=13$ conformations with $\rho=0.73$ (\textit{left}) and $\rho=0.74$
(\textit{right}), which are not the ground-state
conformations but have a just slightly higher energy than
these. The conformations correspond to distinguished
ground-state structures at shorter lengths (helical and
crystallized on the sc lattice, cp. Fig.~\ref{fig:2}). For
reasons of better visibility, the thickness is not shown on
scale.}\label{fig:5}
\end{figure}

It is not surprising, that the situation becomes more
complex with increasing chain length. At least some of the
described ``nice-looking phases'' above are artificial in
the sense, that they occur at exactly one short length, or
are favored just by that very short length, respectively. We
see, for example, for $N=10$ and $N=13$ no exact
($\alpha$-)helices anymore, it rather seems that at these
lengths ``tertiary'' effects already play a role in the
sense, that two small secondary structures are formed which
are then arranged ``side by side''. An indication for this
trend may be that conformations with low thickness are often
``symmetric''\footnote{With ``symmetric'' we mean a somehow
defined symmetry of torsion angles, e.g., torsion angles are
pairwise equal relative to the middle of the chain. It is
generally not essential for defining or distinguishing
different ``phases'', but it is an interesting property and
helps the understanding. Corresponding symmetry observables
are not shown.}, i.e., the conformations get buckled and
turn back at some point (generally in the middle). See for
example the helical region for $N=13$ in
Fig.~\ref{fig:2}. Anyhow, the helical structures being
present for shorter chains indeed exist ``very close'' to
the ground states, i.e., with a slightly higher energy. Two
of these conformations are depicted in Fig.~\ref{fig:5}.

We will get further convincing arguments for this
classification scheme by investigating the thermodynamic
behavior of these polymers in the aforementioned general
structural phases~\cite{partII,letter}. The transition
lines between the phases then depend indeed on both thickness and
temperature. For low temperatures, the helical phase
corresponds to polymers with low thickness, the sheet phase
to a little higher thickness and the ring phase to the very
thick polymers.

\subsection{Selected Problems: Deeper Analysis and Remarks}
\label{sec:selected}

\subsubsection{Simple cubic symmetry}\label{cubes}

One of the basic observations within our
polymer model relates to the fact, that the theory's
parameter space of thickness $\rho$ and $r_{ij}^{\rm min}$
values possibly contains a likewise finite set of isolated
and special parametric points, for which ground-state
conformations exhibit a strict crystalline structure. These
polymer conformations are characterized by a point set of
monomer positions, that is frozen into a regular
three-dimensional (3d) structure and none positional degree
of freedom is left over. Additionally, these ground states
inherit a finite and possibly large ground-state degeneracy
as there exist many ways to arrange the polymers sequence of
monomers onto the frozen point set, without lifting the
theory's energy to larger values. Typically, we then expect a
number of ground-state configurations that increases
exponentially, $n_0
\propto {\rm exp}(cN)$, with the chain length for long
chains. Furthermore, the presence of crystalline ground
states in the tube model possibly is attached to a
triplet of global radius values $r_{\rm gc}=0.5774$, $1/
\sqrt{2}=0.7071$, and $r_{\rm gc}=1$, which denote the radii of
circular polymers, that have an end-to-end distance of unity
with exactly $N=3$, $4$, and $N=6$ monomers on a
circle. With numerical means it is then easy to show, that
these particular radii result into triangular lattice
($N=3$), simple square lattice ($N=4$), and honeycomb
lattice ($N=6$) ground-state polymer point sets in two
dimensions (2d), that is to say dimensional reduced tube
polymer model. The phase space of 3d polymers as such is
much larger than in 2d and a search at $r_{\rm gc}=0.5774$
and $r_{\rm gc}=1$ does not reveal any crystals for 3d
polymers that would persist in the thermodynamic
limit. However, at the particular value of $r_{\rm
gc}=1/\sqrt{2}$ and for $r_{ij}=1$, we find stable 3d
ground-state conformations with point sets of simple cubic
symmetry and with large ground-state degeneracy. These
crystals are likely to extend in the thermodynamic limit for
large $N$ values.

The numerical simulations in case of the 3d thick tube
model have been performed on chain length values
$N=8,9,\ldots,32,36,40,44,48,52,56,60$, and $N=64$. The glob\-al
radius parameter was chosen to be $\rho=1/\sqrt{2}$ with
the value $r_{ij}^{\rm min}=1$ for the position of the
Lennard--Jones potential minimum.\footnote{We also modified a
single interaction term of the Lennard--Jones interaction
in-between the polymers end to end: $V=\infty$ for $r \le 1$
and $V=0$ for $r>1$. This facilitates a perfect arrangement
of the polymers monomers at the start and the end on a
simple cubic lattice, if they prefer to be direct neighbors
in space.} We employed parallel tempering simulations in
the temperature interval $0.01 \le T \le 0.5 $ with a
temperature partition, that ensures acceptance rates around
$0.5$ for parallel tempering swaps in-between neighbors in
the temperature. For complete temperature interval coverage
on a $N=32$ chain a total of $39$ temperature replica was
needed. A single Monte Carlo run consists out of $10^9
\times N$ monomer positional updates and from the ensemble
of configurations the minimum-energy configuration was
stored. A sequence of about $10$ continuation runs for each
chain length $N$ with identical run parameters but with
continued start configurations then yields an ensemble of
about $10$ ground-state estimates, and also the global
ground-state estimate of the simulation. \revis{Finally, an adapted
conjugate-gradient method was applied for refinement.} We
found that the efficiency of the Monte Carlo simulation in
an attempt to populate statistically independent ground
states rapidly degrades for chain length values $N \ge 36$
and therefore the longest chains (except the one at $N=36$)
are excluded from the further analysis.

\begin{figure}[b]
\includegraphics[width=.5\textwidth]{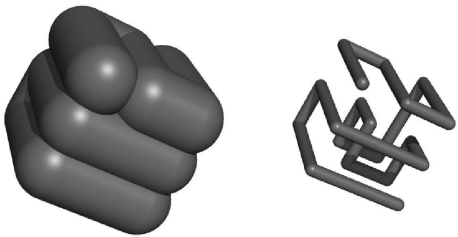}
\caption{A ground-state conformation for the $N=27$ polymer with $\rho=1/\sqrt{2}$ and $r_{ij}^{\rm min}=1$ (i.e., $\sigma\neq1$). The \textit{left} picture shows the conformation in the proper scale, the \textit{right} side shows the same conformation without proper thickness.}  
\label{fig:n27}
\end{figure}

For a crystalline polymer conformation with simple cubic
symmetry one can find a set of transformations, i.e.,
translations and orthochroneous rotations, that map the
polymers point set to some point set of a simple hypercubic
lattice. Denoting by $\vec{x}_i^{\,\rm
sc}=m_i^{\alpha}\vec{e}_{\alpha}$ with $i=1,\ldots,N$,
$\alpha=1,2,3$, and with $m_i^{\alpha}$ integer, a point set
on a simple cubic lattice, the mean squared distance to a
simple cubic lattice
\begin{equation}
d_{\rm sc}^2={1 \over N}\sum_{i=1}^{N} 
( \vec{x}_i^{\rm polymer}-\vec{x}_i^{\,\rm sc})^2
\end{equation}
can be transformed to $d_{\rm sc}^2=0$, if also an
appropriate set of $m_i^{\alpha}$ values is chosen. Our main
numerical result consists in the finding that all of
our ground-state polymers at $N=8,9$ and for
$N=11,12,\ldots,36$ fulfill the numerically determined
inequality $d_{\rm sc}^2 \le 0.000053$ and therefore we
observe a blatant simple hypercubic symmetry in the ground
state of the theory for the considered chain length values
to a high degree of numerical precision. A particular
impressive ground-state conformation is displayed in
Fig. \ref{fig:n27}, where the $N=27$ polymer folds onto a $3
\times 3 \times 3$ cuboid: ${\rm Cub}(3 \times 3 \times
3)$. For the $N=27$ chain, we performed a total of eight
different continuation runs, which all yielded simple cubic
symmetry in the ground-state estimates with values $d_{\rm
sc}^2 \le 0.000053$.  From these eight configurations, six
were found with point sets isomorphic to ${\rm Cub}(3 \times
3 \times 3)$, however with five different mappings of the
polymer sequence to the cuboid and with almost degenerate
energy close to the ground-state energy. Given the numerical
ability of the algorithm to identify different ground-state
and near-by ground-state conformations, it appears unlikely
that the true ground state has not been identified for the
$N=27$ chain. A similar remark applies to all shorter
chains. For purposes of future reference, and as a yard
stick of our numerical precision we display in
\hbox{Table~\ref{tab:sc_energies}} ground-state energy density values
$e_0=E/N$ from numerical simulations (second row) as a
function of the chain length $N$, as well as exactly
calculated energy densities for various cuboids. It is
noteworthy that ground states at $N=8,12$ and $N=18$ also
exhibit cuboidal point sets: ${\rm Cub}(2 \times 2 \times
2), {\rm Cub}(2 \times 2 \times 3)$ and ${\rm Cub}(2 \times
3 \times 3)$ respectively. For the $N=36$ chain the exactly
calculated energy density of the ${\rm Cub}(3 \times 3
\times 4)$ cuboid undershoots the numerical value
significantly and, in fact, the cuboid was not found in the
numerical simulations. This again indicates the failure of
our numerical algorithms for chains with length $N \ge
36$. Finally, it is also of interest to classify the secondary
structures of compactified ground-state conformations for the
case of simple cubic symmetries.  In particular we may
consider U-turns (planar), and simple cubic helices
(3-dimensional)~\cite{dill90jcp}, which both are chain
segments of four monomers with bending angles of
$90$-degrees in-between consecutive segments on the simple
cubic lattice. Using a pattern recognition program along the
ideas of Tenenbaum et al.~\cite{tenenbaum00science} on a
set of five different ground-state conformations for the
$N=27$ polymer we obtain rather low probabilities $P_{\rm
U-turn}
\approx 0.27$ and $P_{\rm sc-helix} \approx 0.22$ for the
occurrence of U-turns and sc-helices, respectively.

\begin{table}
\caption{Ground-state energy density values $e_0=E/N$
from numerical simulations (\textit{second row}) as a
function of the chain length $N$, as well as exactly
calculated energy densities $e_0({\rm cuboids})$ for various
cuboids (\textit{fourth row}).}
\label{tab:sc_energies}
\begin{tabular*}{\columnwidth}{c@{\extracolsep{\fill}}ccc}
\hline\noalign{\smallskip}
$N$ &  $e_0$ & ${\rm Manifold}$ & $e_0({\rm cuboids})$\\
\noalign{\smallskip}\hline\noalign{\smallskip}
        8  & \revis{$-1.0032$} & ${\rm Cub}(2 \times 2 \times 2)$ & $-1.0038$ \\
        9  & \revis{$-0.9670$} & & \\
       10  & \revis{$-1.0398$} & & \\
       11  & \revis{$-1.1052$} & & \\
       12  & \revis{$-1.2497$} & ${\rm Cub}(2 \times 2 \times 3)$ & $-1.2514$ \\
       13  & \revis{$-1.2263$} & & \\
       14  & \revis{$-1.2773$} & & \\
       15  & \revis{$-1.3179$} & & \\
       16  & \revis{$-1.4152$} & & \\
       17  & \revis{$-1.4451$} & & \\
       18  & \revis{$-1.5267$} & ${\rm Cub}(2 \times 3 \times 3)$ & $-1.5293$ \\
       19  & \revis{$-1.5165$} & & \\
       20  & \revis{$-1.5496$} & & \\
       21  & \revis{$-1.5729$} & & \\
       22  & \revis{$-1.6346$} & & \\
       23  & \revis{$-1.6612$} & & \\
       24  & \revis{$-1.7159$} & & \\
       25  & \revis{$-1.7465$} & & \\
       26  & \revis{$-1.7944$} & & \\
       27  & \revis{$-1.8385$} & ${\rm Cub}(3 \times 3 \times 3)$ & $-1.8433$ \\
       28  & \revis{$-1.8230$} & & \\
       29  & \revis{$-1.8282$} & & \\
       30  & \revis{$-1.8383$} & & \\
       31  & \revis{$-1.8750$} & & \\
       32  & \revis{$-1.8875$} & & \\
       36  & \revis{$-1.9224$} & ${\rm Cub}(3 \times 3 \times 4)$ & $-2.0036$ \\
\noalign{\smallskip}\hline
\end{tabular*}
\end{table}

\subsubsection{The $\alpha$-helix region}
\label{sec:alpha}

For the $N=8$ and $9$ polymer, we found a thickness region,
where the $\alpha$-helix is the ground-state conformation
(see Sect.~\ref{sec:A}). Remember that we used the
Lennard--Jones potential with $\sigma=1$ there, which sets
the interaction length scale. There is nothing special with
it, except that the potential just vanishes at the bond
length, a fact that plays just a ``second order'' role, as we
are not counting energy contribution from consecutive
monomers at all.

\begin{figure}[b!]
\includegraphics[width=\textwidth]{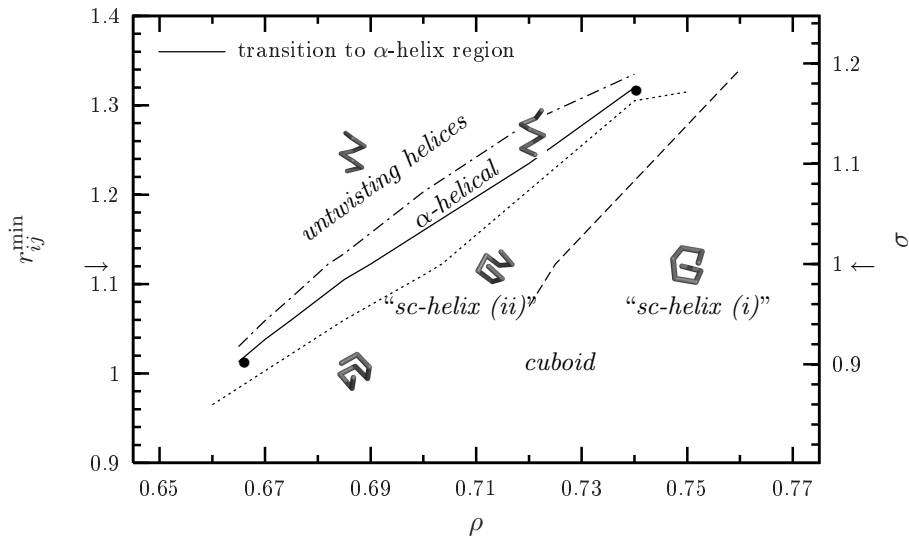}
\caption{$N=8$: The $\sigma$--$\rho$ plane and ground-state conformations near
the $\alpha$-helix. The left and right coordinates are
connected just via $r_{ij}^{\rm min}=2^{1/6}\sigma$. See
text for details.}\label{fig:6}
\end{figure}

Because of the special role of the $\alpha$-helix in nature
(besides its geometrical elegance), we will here try to
track the $\alpha$-helix not only in the thickness but also
in the $\sigma$-direction of the ``phase space'', i.e.~we
vary $\rho$ and $\sigma$ independently in the vicinity of
the assumed ``$\alpha$-region''. Our results are displayed,
exemplarily for $N=8$, in Fig.~\ref{fig:6}. We see, that
the $\alpha$-helix occurs as ground-state conformation in a
small, bounded thickness interval ($0.66<\rho<0.75$) right
``before'' an abrupt conformational change (depicted by the
\textit{solid line}) to cubelike structures. The transition
line increases approximately linearly in the
interaction-length--thickness plane, a dependence, which
seems to hold generally for structural transitions in the
vicinity. Following a perpendicular path, i.e., with
increasing interaction lengths and decreasing thickness, the
helices untwist smoothly. The \textit{dashed-dotted} line
together with the \textit{solid line} define the region,
where the $\alpha$-helix is the ground state of the system
(cp. insets in Fig.~\ref{fig:1}). Note that for $\rho<0.66$
and $\rho>0.75$, $\alpha$-helices are no ground states at
all.\pagebreak

A further interesting transition is marked by the
\textit{dashed line} in Fig.~\ref{fig:6}. This line
indicates the transition bet\-ween the so-called simple cubic
``lattice helices~(ii)'' and ``(i)''~\cite{dill90jcp}, i.e.,
cuboidlike structures with parallel and antiparallel tails
(remember that for $r_{ij}^{\rm min}=1$ and
$\rho=1/\sqrt{2}$, we observe the ``crystallization''
exactly at the simple cubic lattice, as mentioned in
Sect.~\ref{sec:A_thin} and in detail discussed in
Sect.~\ref{cubes}).

For the sake of completeness, the \textit{dotted line}
indicates a conformational change to some less interesting
intermediate structure ``between'' $\alpha$- and lattice
helices and the \textit{arrows} on the $y$-axes mark the
line $\sigma=1$ investigated in the first part of this
study.

\section{\label{sec:summary}Summary}

The aim of this work was to take the simplest coarse-grained
model for off-lattice polymers with explicit volume
exclusion and to show, to which degree polymer
crystallization can be understood even with this simplest
model. We introduced the polymer volume using the concept of
the radius of curvature which is indeed, in the first
instance, a mathematical concept. In fact, it has been
proven that ``[it] is connected to various physically
appealing properties of a curve. In particular, [it]
provides a concise characterization of the thickness of a
curve, [\ldots] as have been investigated within the context
of DNA''~\cite{gonzmad99pnas}, it was further successfully
used in more complex models for
proteins~\cite{banamarit03jpcm,banamarit03rmp,banagonz03jsp}
and it was finally shown, that this concept is effectively
equal to a volume exclusion using two-point functions for
polymers in good solvents~\cite{neuh06}.

Using sophisticated simulation techniques, we have analyzed
systematically and in detail ground-state structures for
the described model with fixed interaction length. We have
shown, for example, that already in this simplest model
basic secondary structures like helices and sheets
form. This statement is, due to the simplicity of the
model, valid for various classes of polymers. Of course, it
should be stated as well, that mentioned structures are not
very stable against variations of the thickness, but this
was not expected either.

We investigated furthermore in detail the ``neighborhood''
of the $\alpha$-helix by varying both, thickness and
interaction length. Affirming above statement, it turned
out that the $\alpha$-helix exists as ground state only in
a small, bounded area in the $\sigma$--$\rho$ space, but is
surrounded by other helical and helicallike conforma\/tions.

It was of course known for a long time, that helices and
sheets form within coarse-grained models including a somehow
defined volume exclusion, but to our knowledge mainly
for dedicated or less simple and not that
general models. In some interesting works, for example,
\revis{the
strength of directionalized interactions~\cite{kemp98prl},}
explicit hydrogen bonds~\cite{bana04pnas,wolff08gene},
solvent
particles~\cite{snir05sci,snir07pre,hansgo07prl}\revis{, or
the interplay between attractive interactions and
packing~\cite{magee06prl} play a role. In particular the
findings of~\cite{magee06prl} also confirm the existence of
not too long helical structures in a specific homopolymer
model that is characterized by strong repulsive interactions
between spheres.} It is a common ansatz to
investigate and understand protein folding, stressing that
we do not speak only of proteins but of a general class of
polymers including proteins, at different abstraction
(coarse grained) levels. It seems that at least parts of the
general secondary structure formation can be attributed
already to the simplest generic model for thick
polymers. \revis{Obviously, these secondary structure
segments have to be strengthened by further
interactions in order to reach, for example, biologically
relevant structure sizes, as tertiary effects set in at
comparatively short chain lengths in simple coarse-grained
models~\cite{kemp98prl,snir07pre,partII}.}

The analysis of ground states is, of course, just a first
step to an understanding of the model. In a subsequent
work~\cite{partII} we will, based on the knowledge of the
ground-state conformations, focus on the thermodynamic
behavior and conformational phases at finite temperatures.

\paragraph{Acknowledgement}
This work is partially supported by the DFG (German
Science Foundation) under Grant No. JA 483/24-1/2/3 and
the Graduate School of Excellence ``BuildMoNa''. Some
simulations were performed on the supercomputer JUMP
of the John von Neumann Institute for Computing (NIC),
Forschungszentrum J\"ulich, under Grant No. hlz11.

\end{document}